\def\sorb{b}
\if s\sorb
  \documentstyle{article}
  \newlength{\absize}
  \renewcommand{\baselinestretch}{1.5}
  \renewcommand{\arraystretch}{0.5}
  
\begin{document}
  \date{}
  \pagestyle{empty}
  \thispagestyle{empty}
  \renewcommand{\thefootnote}{\fnsymbol{footnote}}
  \newcommand{\starttext}{\newpage\normalsize
    \pagestyle{plain}
    \setlength{\baselineskip}{4ex}\par
    \twocolumn\setcounter{footnote}{0}
    \renewcommand{\thefootnote}{\arabic{footnote}}}
\else
  \documentstyle[12pt,a4wide,epsf]{article}
  \newlength{\absize}
  \setlength{\absize}{\textwidth}
  \renewcommand{\baselinestretch}{2.0}
  \renewcommand{\arraystretch}{0.5}
  \begin{document}
  \thispagestyle{empty}
  \pagestyle{empty}
  \renewcommand{\thefootnote}{\fnsymbol{footnote}}
  \newcommand{\starttext}{\newpage\normalsize
    \pagestyle{plain}
    \setlength{\baselineskip}{4ex}\par
    \setcounter{footnote}{0}
    \renewcommand{\thefootnote}{\arabic{footnote}}}
\fi
\newcommand{\preprint}[1]{%
  \begin{flushright}
    \setlength{\baselineskip}{3ex} #1
  \end{flushright}}
\renewcommand{\title}[1]{%
  \begin{center}
    \LARGE #1
  \end{center}\par}
\renewcommand{\author}[1]{%
  \vspace{2ex}
  {\Large
   \begin{center}
     \setlength{\baselineskip}{3ex} #1 \par
   \end{center}}}
\renewcommand{\thanks}[1]{\footnote{#1}}
\renewcommand{\abstract}[1]{%
  \vspace{2ex}
  \normalsize
  \begin{center}
    \centerline{\bf Abstract}\par
    \vspace{2ex}
    \parbox{\absize}{#1\setlength{\baselineskip}{2.5ex}\par}
  \end{center}}

\setlength{\parindent}{3em}
\setlength{\footnotesep}{.6\baselineskip}
\newcommand{\myfoot}[1]{%
  \footnote{\setlength{\baselineskip}{.75\baselineskip}#1}}
\renewcommand{\thepage}{\arabic{page}}
\setcounter{bottomnumber}{2}
\setcounter{topnumber}{3}
\setcounter{totalnumber}{4}
\newcommand{\figsize}{}
\renewcommand{\bottomfraction}{1}
\renewcommand{\topfraction}{1}
\renewcommand{\textfraction}{0}
\newcommand{\beq}{\begin{equation}}
\newcommand{\eeq}{\end{equation}}
\newcommand{\beqa}{\begin{eqnarray}}
\newcommand{\eeqa}{\end{eqnarray}}
\newcommand{\nn}{\nonumber}

\newcommand{\dd}{\mbox{{\rm d}}}
\newcommand{\mH}{m_{\rm H}}
\newcommand{\dLips}{\mbox{{\rm dLips}}}

\def\Im{\mbox{\rm Im\ }}
\def\fourth{\textstyle{1\over4}}
\def\gsim{\mathrel{\rlap{\raise 1.5pt \hbox{$>$}}\lower 3.5pt
\hbox{$\sim$}}}
%
\def\slash#1{#1 \hskip -0.5em /}
%
%
           
\preprint{University of Bergen, Department of Physics \\ 
Scientific/Technical Report No.\ 1994-06 \\ ISSN~0803-2696\\
February 1994}

\vfill
\title{Signals of $CP$ violation in Higgs decay}

\vfill
\author{Arild Skjold \\ Per Osland \\\hfil\\
        Department of Physics\thanks{Electronic mail addresses:
                {\tt \{skjold,osland\}@vsfys1.fi.uib.no}}\\
        University of Bergen \\ All\'egt.~55, N-5007 Bergen, Norway }
\date{}

\vfill
\abstract{We consider an extension of the Standard Model where some
Higgs particle is not an eigenstate of $CP$, and discuss the possibility of
extracting signals of the resulting $CP$ violation. 
In the case of Higgs decay to four fermions we study
correlations among  momenta of the final-state fermions. 
We discuss observables which may demonstrate presence of $CP$ violation
and identify a phase shift $\delta$, which is a measure of the strength
of $CP$ violation in the Higgs-vector-vector coupling,
and which can be measured directly in the decay distribution.
In addition to these angular correlations, we
consider correlations between energy differences. The former correlations
include some recently reported results, whereas the latter ones
appear to provide a much better probe for revealing $CP$ violation.}
  
\vfill

\starttext


The origin of mass and the origin of $CP$ violation are widely considered
to be the most fundamental issues in contemporary particle physics.
Perhaps for this very reason, there has been much speculation
on a possible connection.  This is also in part caused by 
the fact that the amount of $CP$ violation accommodated in the CKM matrix
does not appear sufficient to explain the observed baryon
to photon ratio \cite{Cohen}.

We shall here consider the possibility that $CP$ violation is present 
in the Higgs sector, and discuss some possible signals of such effects
in decays of Higgs particles.
While the standard model induces $CP$ violation in the Higgs sector at 
the one-loop level provided 
the Yukawa couplings contain both scalar and pseudoscalar components
\cite{Wein}, we actually have in mind 
an extended model, such as e.g., the two-Higgs-doublet model
\cite{TDLee}.

Below we postulate an effective Lagrangian which contains $CP$ violation in
the Higgs sector. However, when the source of $CP$ violation is turned off, 
the interaction reduces to that of the Standard Model. 
In cases considered in the literature, $CP$ violation usually 
appears as a one-loop effect. This is due to the fact that the $CP$ odd 
coupling introduced below
is a higher-dimensional operator and in renormalizable models these are
induced only at loop level. Consequently we expect the
effects to be small and any observation of $CP$ violation to be 
equally difficult.

$CP$ non-conservation has manifested itself so far only in the neutral kaon
system. In the context of the Standard Model this $CP$ violation 
originates from the Yukawa sector via the CKM matrix \cite{KoMa}. Although 
there may be several sources of $CP$ violation, including the one above, we 
will here consider a simple model where the $CP$ violation is restricted
to the Higgs sector and in particular to the coupling 
between some Higgs boson and the vector bosons. 
Specifically, by assuming that the coupling between the Higgs boson $H$ 
and the vector bosons $V=W, Z$ has both scalar 
and pseudoscalar components, 
the effective Lagrangian for the $HVV$-vertex may be written as
\cite{Nel,GrGu,Cha}
\beq
{\cal L}_{HVV} = 2\cdot 2^{1/4} \sqrt{G_{\rm F}} \left[ 
m^{2}_V \ V^\mu V_\mu \ H
+ \fourth\, \eta \ \epsilon^{\mu \nu \rho \sigma}\
V_{\mu \nu} V_{\rho \sigma} \ H\right],
\label{EQU:int1}
\eeq
with $G_{\rm F}$ the Fermi constant. 
Furthermore, $V_{\mu \nu}=\partial_\mu V_\nu -\partial_\nu V_\mu$.
The parameter $\eta$ is, according to the discussion above,
a small dimensionless quantity that depends on the kinematics.
However, we will keep $\eta$ arbitrary and our results are valid
for any value of $\eta$. 
The first term in ${\cal L}_{HVV}$ is $CP$ even, whereas 
the second one is $CP$ odd.
Simultaneous presence of both terms leads to $CP$ violation. 
However, we assume that $CPT$ is conserved.
This implies that $\eta$ must be real.

We shall consider the decay of a Higgs via two 
vector bosons ($W^+W^-$ or $ZZ$), to two 
non-identical fermion-antifermion pairs,
$H \rightarrow V_1V_2\rightarrow (f_1\bar f_2)(f_3\bar f_4)$.
Let the momenta of the two fermion-antifermion pairs 
($q_1$, $q_2$, $q_3$, and $q_4$ in the Higgs rest frame) 
define two planes, 
and denote by $\phi$ the angle between those two planes
(see eq.~(\ref{EQU:Dj4}) below). 
Then we shall discuss the angular distribution of the decay rate
$\Gamma$,
\beq
\frac{1}{\Gamma}\:
\frac{\dd\Gamma}{\dd\phi} 
\label{EQU:intro1}
\eeq
and an energy-weighted decay rate, to be defined below.

Related studies have been reported by \cite{Nel,Cha}. 
The present discussion will be more general 
and in addition we consider correlations between energy differences. 
Under suitable experimental conditions these energy correlations 
provide a better signal of $CP$ violation.

The fermion-vector coupling is given by
$$
- \frac{i}{2 \sqrt{2}} \gamma^\mu(g_V-g_A\gamma_5), 
$$
where $g_V$ and $g_A$ denote the vector and axial-vector parts 
of the couplings.
As a para\-meterization of these, we define
the angles $\chi_1$ and $\chi_2$ by
\beq
g^{(i)}_{V}  \equiv  g_i \cos \chi_i, \qquad
g^{(i)}_{A} \equiv g_i \sin \chi_i, \qquad i=1,2.
\label{EQU:Dj9} 
\eeq
The only reference to these angles is through $\sin2\chi_i$. 
Relevant values are given in table~1 of ref.~\cite{osskj}.
From (\ref{EQU:int1}), the coupling of $H$ to the vector bosons is given by
\beq
i\ \left(2\cdot 2^{1/4}\right) \sqrt{G_{\rm F}} 
\left[ m_V^2 \ g^{\mu \nu} 
+ \eta\left(k_1^2,k_2^2\right) 
\ \epsilon^{\mu \nu \rho \sigma} k_{1 \rho} k_{2 \sigma} \right],
\label{EQU:int2}
\eeq
with $k_{j}$ the momentum of vector boson $j$, $j=1,2$.
 
The decay rate can be written as
\beq
\dd^8\Gamma
= \sqrt{2}\, \frac{G_{\rm F}}{m}\, 
N_1 N_2 D\biggl[X + {\eta\over m_V^2}\, Y 
+ \left({\eta\over m_V^2}\right)^{2} \, Z \biggr] 
\dLips(m^2;q_1,q_2,q_3,q_4) ,
\label{EQU:utgpkt} 
\eeq
with $m$ the Higgs mass and 
$\dLips(m^2;q_1,q_2,q_3,q_4)$ denoting the Lorentz-invariant phase 
space. Furthermore, $N_j$ is a colour factor, which is three for 
quarks, and one for leptons. 
The momentum correlations are in the massless fermion approximation
given by
\beqa
X
& = & X_{+} + \sin 2 \chi_{1} \sin 2 \chi_{2} \ X_{-}, \nn \\
Y
& = & -\epsilon_{\alpha\beta\gamma\delta}\,
q_{1}^{\alpha}q_{2}^{\beta}q_{3}^{\gamma}q_{4}^{\delta} \
\left[
Y_{-} + \sin(2\chi_{1}) \sin(2\chi_{2})Y_{+} \right], \nn \\
Z
& = & -2 X_{-}^{2}+ \frac{1}{4}\, s_1 s_2 \bigl[Z_1
+ \sin(2\chi_{1}) \sin(2\chi_{2}) Z_2 \bigr],
\eeqa
where
\beqa
X_{\pm}
& = &
(q_1\cdot q_3)(q_2\cdot q_4) \pm(q_1\cdot q_4)(q_2\cdot q_3), \nn \\
Y_{\mp}
& = &
(q_1\mp q_2)\cdot(q_3\mp q_4), \nn \\
Z_1
& = &
  [(q_1\cdot q_3)+(q_2\cdot q_4)]^2
+ [(q_1\cdot q_4)+(q_2\cdot q_3)]^2-\frac{1}{2}\, s_1 s_2, \nn \\
Z_2
& = &
[(q_1+q_2)\cdot(q_3-q_4)][(q_1-q_2)\cdot(q_3+q_4)].
\label{EQU:eqYA}
\eeqa
The normalization in eq.~(\ref{EQU:utgpkt}) involves the function 
\beq
D(s_1,s_2) = m_V^4 \,\prod_{j=1}^{2}
\frac{g_j^2}{(s_j-m_V^2)^2+m_V^2\Gamma_V^2}\:, 
\label{EQU:defc} 
\eeq
with
$$s_1 \equiv (q_1+q_2)^2,   \qquad
s_2 \equiv (q_3+q_4)^2.  $$
Finally, $m_V$ and $\Gamma_V$ denote the mass and total width of
the relevant vector boson, respectively. 

We first consider angular correlations.
The relative orientation of the two planes is defined by the angle
$\phi$,
\beq
\cos \phi = \frac{\left({\bf {q}_{1}} \times {\bf {q}_{2}}\right)
            \cdot \left({\bf {q}_{3}} \times {\bf {q}_{4}}\right)}
                      {|{\bf {q}_{1}} \times {\bf {q}_{2}}| 
                       |{\bf {q}_{3}} \times {\bf {q}_{4}}|}.
\label{EQU:Dj4}
\eeq
We find
\beqa
 \frac{\dd^3\Gamma }{\dd\phi\: \dd s_1 \dd s_2}
& = &\frac{\sqrt{2}}{72 (4\pi)^6}\, N_1 N_2 \, \frac{G_{\rm F}}{m^3}
\sqrt{\lambda\left(m^2,s_1,s_2\right)}\, D(s_1,s_2) \nn \\
& & \times 
\left[X^{\prime} +
{\eta\over m_V^2}\sqrt{\lambda\left(m^{2},s_{1},s_{2}\right)} \ Y^{\prime}
+\left({\eta\over m_V^2}\right)^2\, \lambda\left(m^{2},s_{1},s_{2}\right) \ 
Z^{\prime} \right], 
\label{EQU:no1} 
\eeqa
with 
\beqa
X^{\prime}
& = &\lambda\left(m^{2},s_{1},s_{2}\right)+12 s_{1} s_{2} 
+2s_1 s_2 \cos 2 \phi \nn \\
& & -  \sin 2 \chi_{1} \sin 2 \chi_{2} \left(\frac{3 \pi}{4}\right)^2 
\sqrt{s_1 s_2}\, (m^2-s_1-s_2) \cos \phi,  \nn \\
Y^{\prime}
& = & 
2s_1 s_2 \sin 2 \phi
- \sin 2 \chi_{1} \sin 2 \chi_{2} \,
\frac{1}{2}
\left(\frac{3 \pi}{4}\right)^2 
\sqrt{s_1 s_2}\, (m^2-s_1-s_2) \sin \phi, \nn \\
Z^{\prime}
& = &
2s_1 s_2 \left(1-\frac{1}{4}\, \cos 2 \phi \right),
\label{EQU:no2} 
\eeqa
and where 
$\lambda\left(x,y,z\right)\equiv
x^{2}+y^{2}+z^{2}-2\left(x y + x z + y z\right)$
is the usual two-body phase space function. 

The distribution (\ref{EQU:no1}) can be written in a more compact form as
\beqa
 \frac{\dd^3\Gamma }{\dd\phi\: \dd s_1 \dd s_2}
& = &\frac{\sqrt{2}}{72 (4\pi)^6}\, N_1 N_2 \, \frac{G_{\rm F}}{m^3}
\sqrt{\lambda\left(m^2,s_1,s_2\right)}\, D(s_1,s_2)\  \nn \\
& & \times
\biggl[ \lambda\left(m^2,s_1,s_2\right)+4 s_1 s_2 \left(1+2 \rho^{2} \right) 
+2s_1 s_2\,\rho^2\,\cos 2(\phi - \delta) \nn \\
& & -  \sin 2 \chi_{1} \sin 2 \chi_{2} \left(\frac{3 \pi}{4}\right)^2
\sqrt{s_1 s_2}\, (m^2-s_1-s_2)\,\rho\,\cos (\phi - \delta) \biggr], 
\label{EQU:no3}
\eeqa
with a modulation function
\beq
\rho=\sqrt{1+\eta^2\lambda\left(m^2,s_1,s_2\right)/(4m_V^4)}, 
\eeq
and an angle 
\beq
\delta=
\arctan\frac{\eta(s_1,s_2)\sqrt{\lambda(m^2,s_1,s_2)}}{2m_V^2}, \qquad
-\pi/2 < \delta < \pi/2,
\label{EQU:delta}
\eeq
describing the relative shift in the spatial distribution of 
the two decay planes due to $CP$ violation.
We note that this rotation vanishes at the threshold for producing
vector bosons (where $\lambda=0$) and grows with increasing values
of the Higgs mass. 

This relation (\ref{EQU:delta}) can be inverted to give for the 
$CP$-odd term in the coupling:
\beq
\eta=\frac{2m_V^2}{\sqrt{\lambda(m^2,s_1,s_2)}}\, \tan\delta.
\eeq

A more inclusive distribution is obtained if we integrate
over the invariant masses of the two pairs. 
Thus, let us consider
\beq
\frac{\dd\Gamma}{\dd\phi}
= \int_{0}^{m^2}\dd s_1
\int_{0}^{\left(m -\sqrt{s_1}\right)^2}\dd s_2 \:
\frac{\dd^3\Gamma}{\dd\phi \: \dd s_1 \: \dd s_2}. \label{EQU:Dk3}
\eeq
Due to our ignorance concerning 
$\eta=\eta(s_{1},s_{2})$, we have to perform 
the integration over $s_1$ and $s_2$ in the narrow-width approximation. 
This is of course only meaningful above threshold for producing real
vector bosons.
We introduce the ratio $\mu = \left(2m_V/m\right)^2 < 1$.
The distribution of eq.~(\ref{EQU:intro1}) then takes the compact form
\beq
\frac{2 \pi}{\Gamma}\:\frac{\dd\Gamma}{\dd\phi} 
= 1 + \alpha(m) \, \rho^{2} \, \cos 2\left(\phi -\delta\right)
+ \beta(m) \, \rho \, \cos (\phi -\delta),
\label{EQU:Dl5} 
\eeq
with
\beqa
\alpha(m)
&=&
\frac{1}{2} \, \frac{\mu^2}{4(1-\mu)(1+2 \eta^{2})+3\mu^2}, \nn
\\
\beta(m)
&=&
-\frac{1}{2} \sin2\chi_1\, \sin2\chi_2\left(\frac{3\pi}{4}\right)^2\,
\frac{\mu(2-\mu)}{4(1-\mu)(1+2 \eta^{2})+3\mu^2}, 
\eeqa
and
\beq
\delta=\arctan\frac{2\eta\sqrt{1-\mu}}{\mu}, \qquad
\rho =\sqrt{1+4\eta^2\,\frac{1-\mu}{\mu^2}}.
\label{EQU:delro}
\eeq

Measurement of this rotation $\delta$ of the azimuthal distributions,
would demonstrate $CP$ violation in the coupling between the Higgs 
boson and the vector bosons.
In order to facilitate such a possibility, we introduce the following
measures of the asymmetry:
\beqa
\left.
\begin{array}{c}
A\left(m\right) \\
A^{\prime}\left(m\right)
\end{array}
\right\}
& \equiv &
\frac{1}{\pi} \int_{0}^{2\pi}\dd\phi 
\left\{
\begin{array}{c}
\cos 2\phi \\
\sin 2\phi
\end{array}
\right.
\left(\frac{2 \pi}{\Gamma}\:\frac{\dd\Gamma}{\dd\phi}\right) = 
\rho^{2} \ \alpha\left(m\right)
\left\{
\begin{array}{c}
\cos 2\delta \\
\sin 2\delta
\end{array} 
\right.,
\label{EQU:nr1} \\
\left.
\begin{array}{c}
B\left(m\right) \\
B^{\prime}\left(m\right)
\end{array}
\right\}
& \equiv &
\frac{1}{\pi} \int_{0}^{2\pi}\dd\phi 
\left\{
\begin{array}{c}
\cos \phi \\
\sin \phi
\end{array}
\right.
\left(\frac{2 \pi}{\Gamma}\:\frac{\dd\Gamma}{\dd\phi}\right) = 
\rho \ \beta\left(m\right)
\left\{
\begin{array}{c}
\cos \delta \\
\sin \delta
\end{array}
\right., 
\label{EQU:nr2} 
\eeqa
where the unprimed observables
correspond to the SM prediction (modulo corrections of
order $\eta^{2}$) 
and the primed ones correspond to the $CP$ violating contributions. 
This identification is evident when we note that 
(see eqs.~(\ref{EQU:no1}) and (\ref{EQU:no2}))
\beq
\frac{2 \pi}{\Gamma}\:\frac{\dd\Gamma}{\dd\phi} =
1+A\left(m\right) \cos 2\phi + B\left(m\right) \cos \phi 
 +A^{\prime}\left(m\right) \sin 2\phi + B^{\prime}\left(m\right) \sin \phi
\label{EQU:nr5}
\eeq
for any $\eta$.
The introduction of the above asymmetries requires that we 
are experimentally
in a position where we can orient $\phi$ from 0 to $2\pi$. (Of course,
complete jet identification is required for this kind of analysis.
We hope to present Monte Carlo data on the expected efficiency elsewhere.)

Presence of $CP$ violation will now manifest itself as a non-zero value
for the observables $A^{\prime}$ and $B^{\prime}$; the magnitude of 
$CP$ violation may be determined from the angle $\delta$,
\beq
\tan 2\delta = \frac{A^{\prime}\left(m\right)}{A\left(m\right)} \qquad
\mbox{\rm and} \qquad
\tan \delta  = \frac{B^{\prime}\left(m\right)}{B\left(m\right)}.
\label{EQU:nr6}
\eeq 
In principle, this allows for consistency checks, since the two expressions in
eq.~(\ref{EQU:nr6}) are closely related. However, a possible experimental
observation of $CP$ violation is strongly dependent upon the magnitudes
of the observables $A^{\prime}$ and $B^{\prime}$.
A first question in this direction is: which of them is easier to detect?
In order to answer this question, we may study the ratio
\beq
\frac{B^{\prime}\left(m\right)}{A^{\prime}\left(m\right)} =
\frac{\beta\left(m\right)}{2 \alpha\left(m\right)}
\label{EQU:forh1}
\eeq
(we have here made use of the fact that $\rho \cos \delta=1$).
This ratio is {\it independent of $\eta$}.
In the case of $H \rightarrow W^{+}W^{-}\rightarrow 4 f$, 
$B^{\prime}$ is relatively important for any Higgs mass. In the case of 
$H \rightarrow ZZ\rightarrow 4 l$, $A^{\prime}$ is relatively 
important for intermediate Higgs bosons, whereas $B^{\prime}$ becomes
more important for $m\gsim$~600~GeV.
(The case e.g. $H \rightarrow ZZ\rightarrow 2l\,2q$ is intermediate,
cf.\ table~1 in \cite{osskj}.)

\begin{figure}
\refstepcounter{figure}
\label{plfun2}
\begin{center}
\mbox{\epsffile{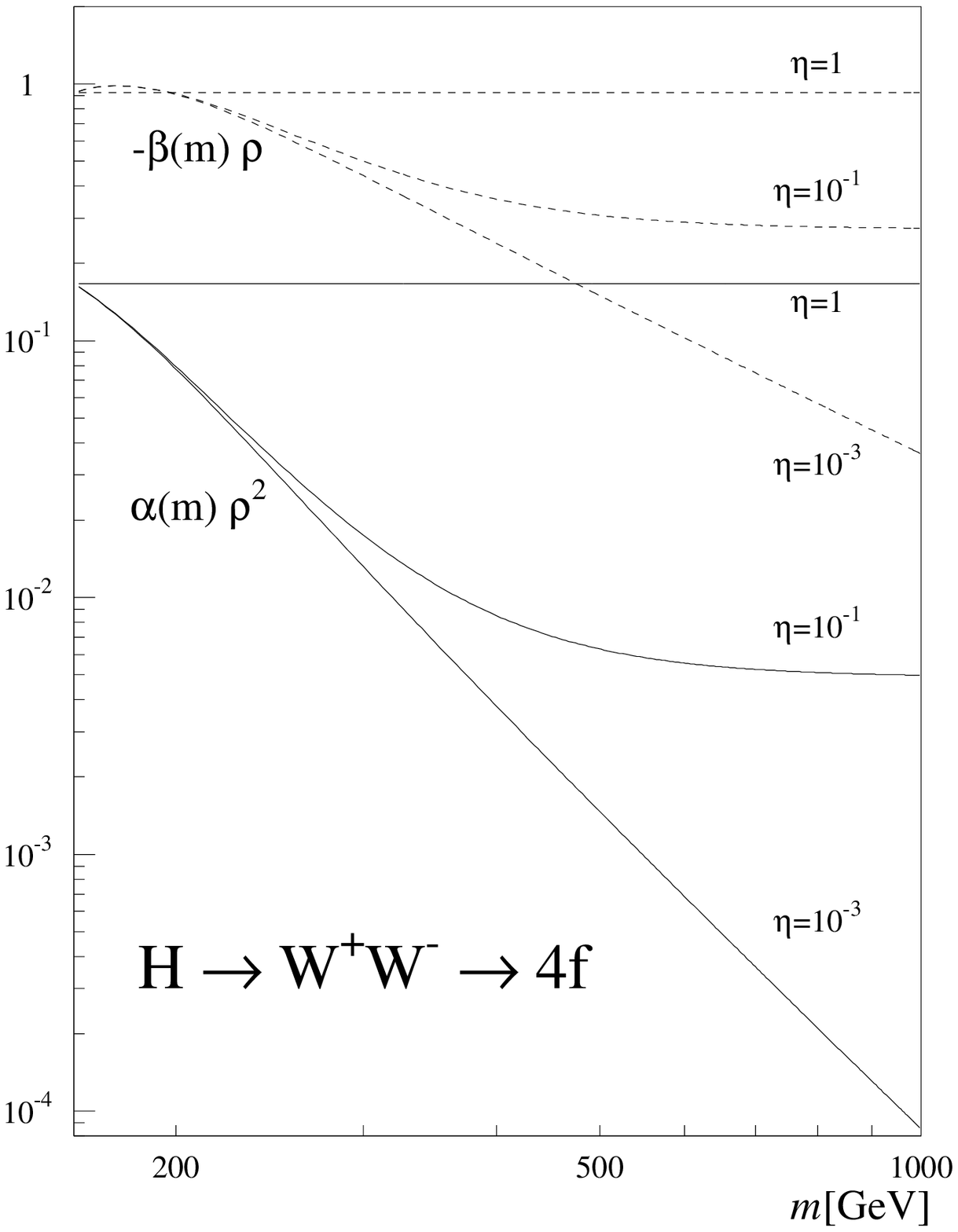}}

\vspace{0mm}

Figure~\thefigure
\end{center}
\end{figure}

\begin{figure}
\refstepcounter{figure}
\label{plfun3}
\begin{center}
\mbox{\epsffile{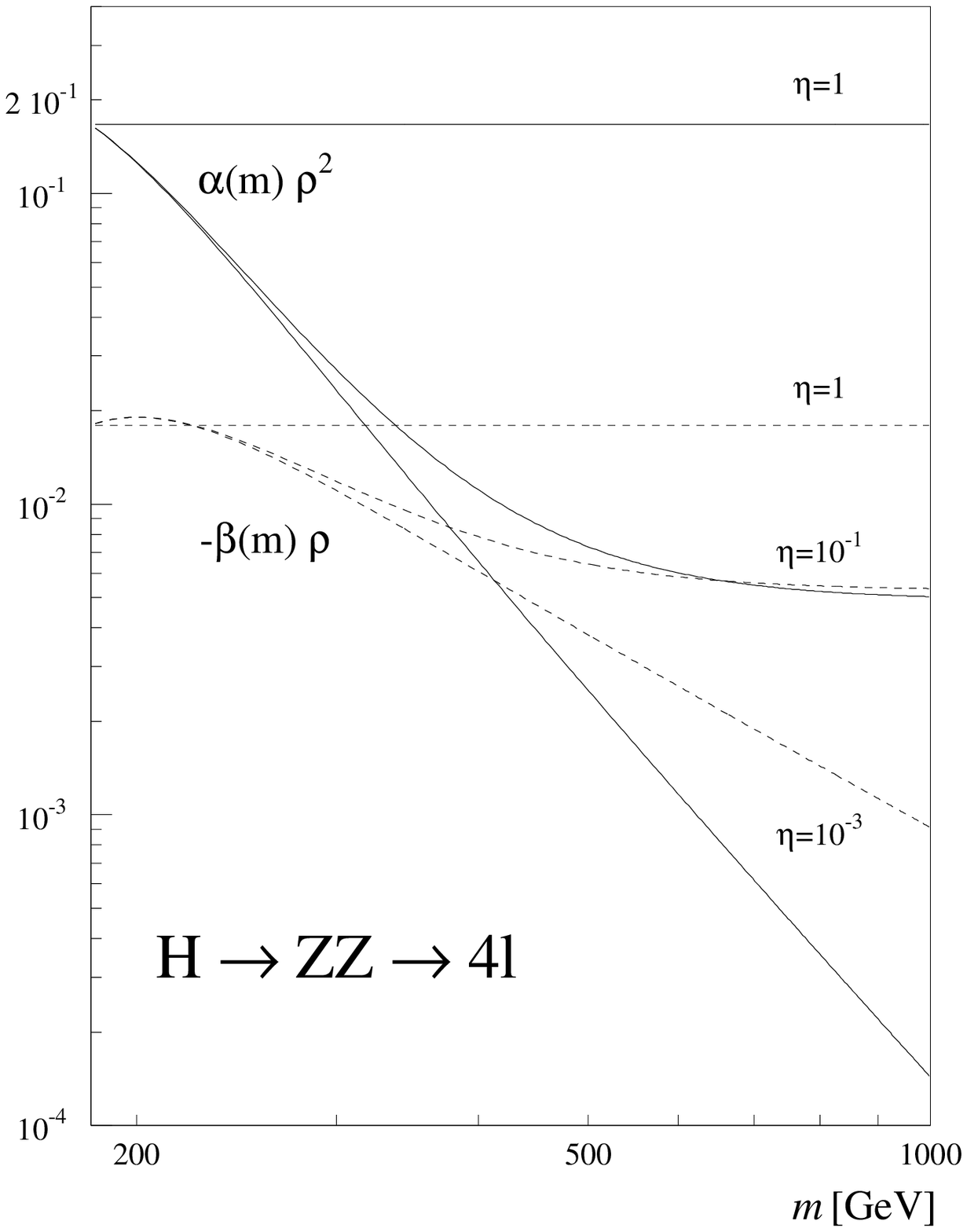}}

\vspace{0mm}

Figure~\thefigure
\end{center}
\end{figure}

The amplitude functions $\alpha(m)\rho^2$ and
$\beta(m)\rho$ of (\ref{EQU:Dl5}) are given in 
figs.~\ref{plfun2}--\ref{plfun3} for the cases 
$H \rightarrow W^{+}W^{-}\rightarrow 4 f$ and 
$H \rightarrow ZZ\rightarrow 4 l$, respectively. 
We note that $\alpha(m)\rho^2$ is independent of the relative
strengths of axial and vector couplings, whereas $\beta(m)\rho$ 
is proportional to the $\sin2\chi_i$ factors.
Both amplitudes are small in most of the 
$\left(\eta, m, \sin 2\chi_{i}\right)$ parameter-space,
but $\beta(m)\rho$ is comparable to unity
for any value of $\eta$,
in the intermediate Higgs mass range in the case 
$H \rightarrow W^{+}W^{-}\rightarrow 4 f$. 
The two amplitudes are {\it independent} of the Higgs mass 
{\it iff} $\eta=\pm 1$. 

\begin{figure}
\refstepcounter{figure}
\label{plfun1}
\begin{center}
\mbox{\epsffile{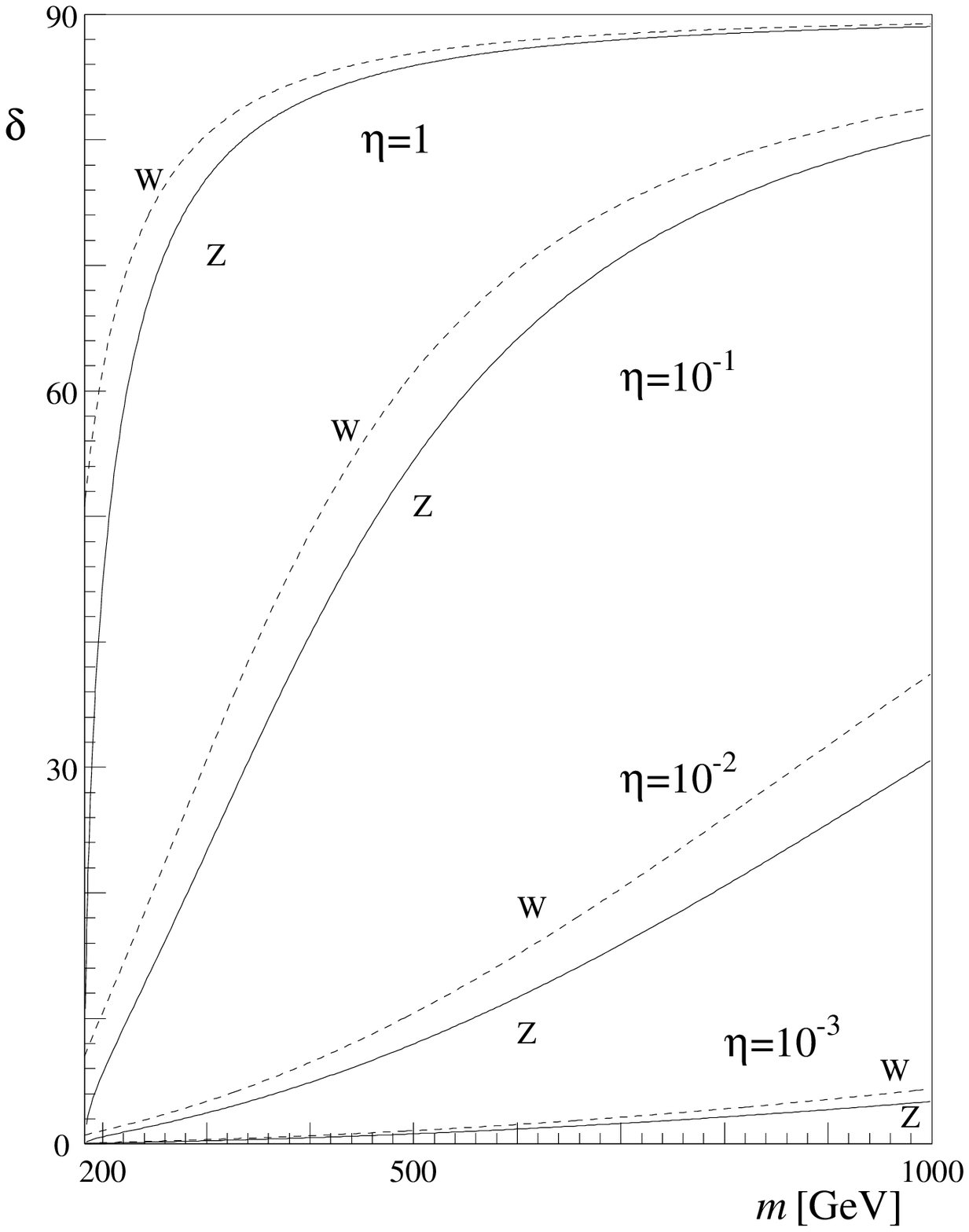}}

\vspace{0mm}

Figure~\thefigure
\end{center}
\end{figure}

The angle $\delta$ is given in fig.~\ref{plfun1} 
for different values of $\eta$. 
Provided $\eta$ is not too small,
this angle is significantly different
from zero when the Higgs is well above threshold.
It appears from the previous figures that an experimental observation 
of $CP$ violation would only be possible in restricted ranges of
$\eta$ and Higgs mass values, and only for selected decay channels.

\begin{figure}
\refstepcounter{figure}
\label{plfun6}
\begin{center}
\mbox{\epsffile{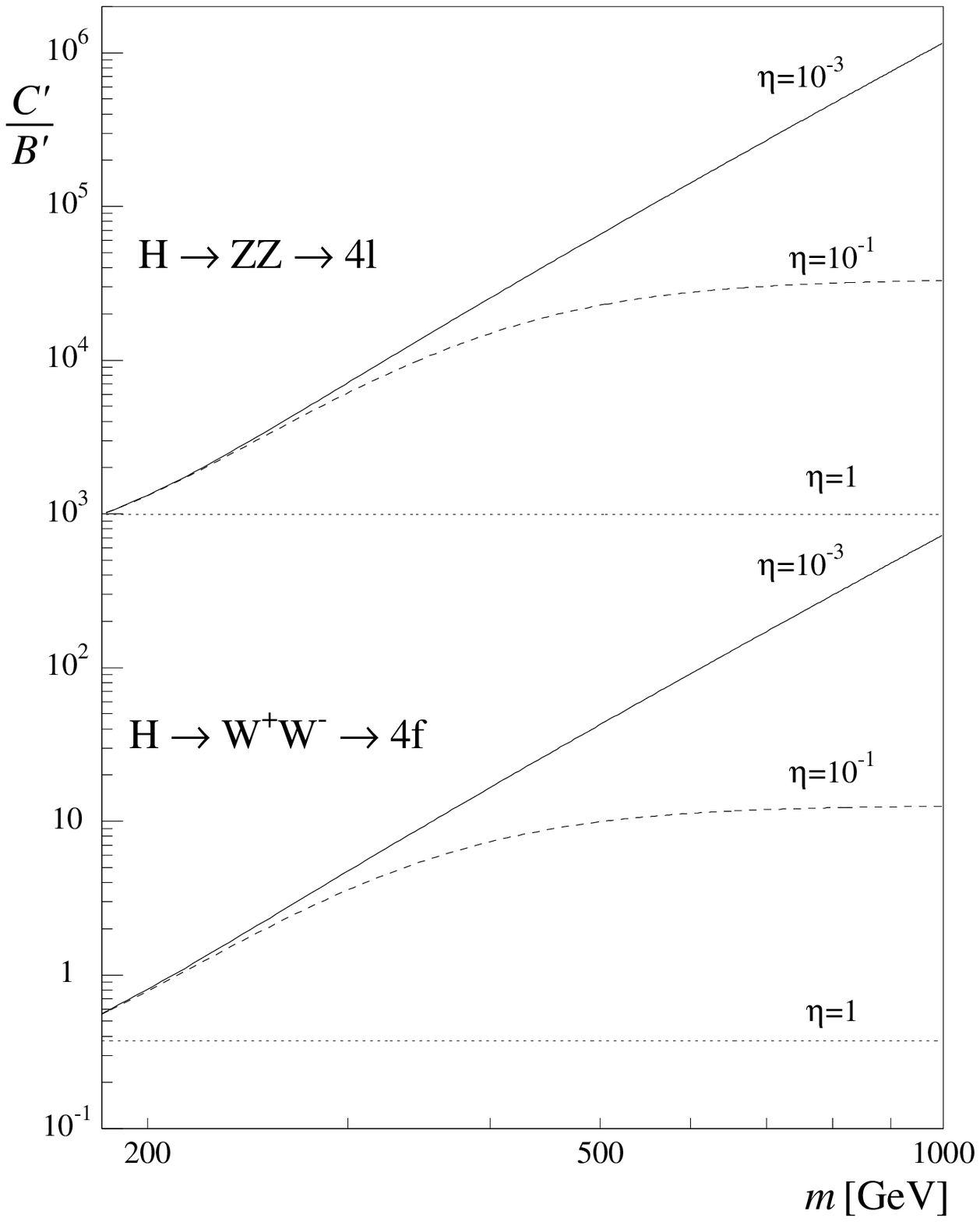}}

\vspace{0mm}

Figure~\thefigure
\end{center}
\end{figure}

\begin{figure}
\refstepcounter{figure}
\label{plfun4}
\begin{center}
\mbox{\epsffile{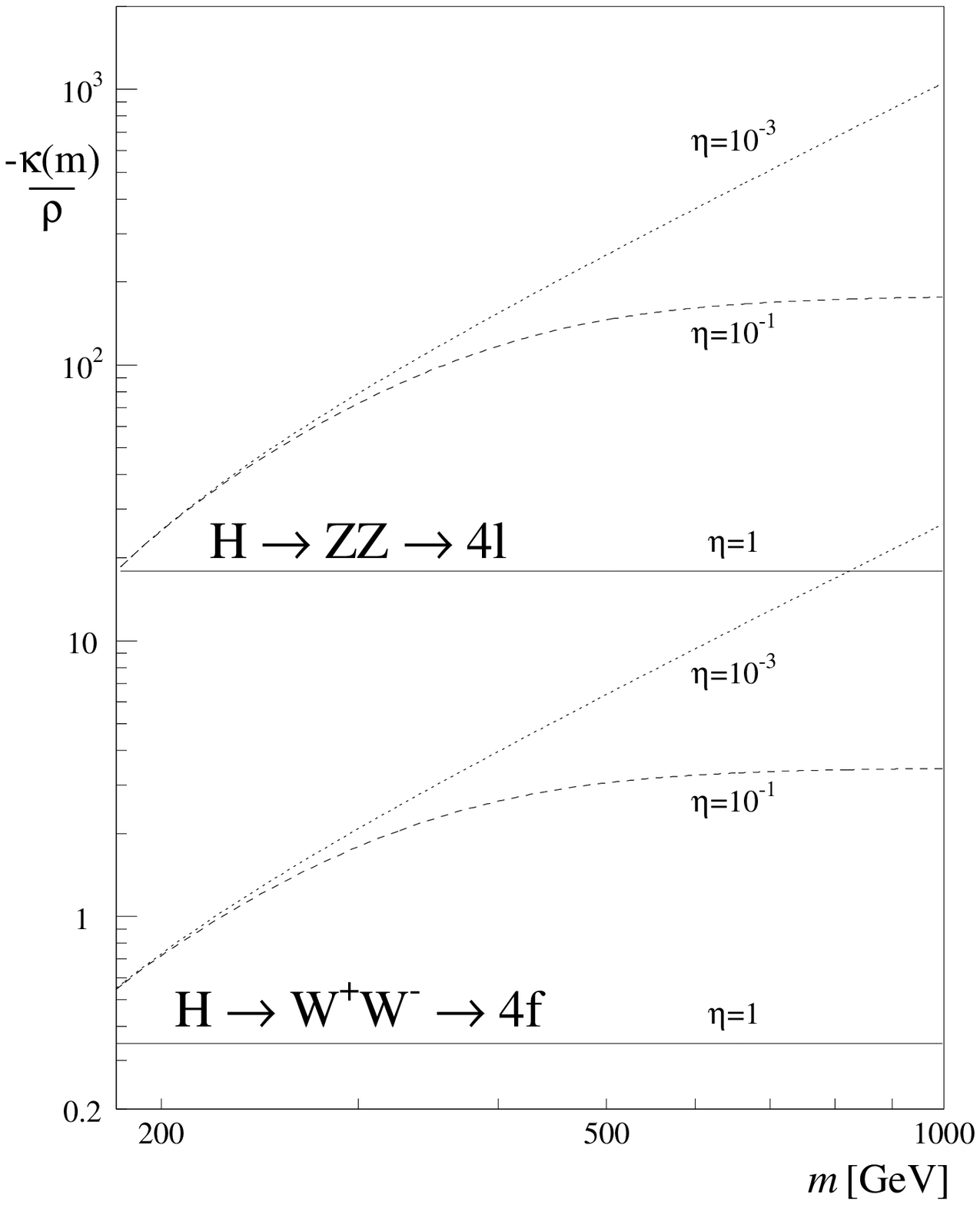}}

\vspace{0mm}

Figure~\thefigure
\end{center}
\end{figure}

Let us now turn to a discussion of the decay rate weighted
with energy differences. 
We multiply (\ref{EQU:utgpkt}) by the energy differences 
$(\omega_{1}-\omega_{2})(\omega_{3}-\omega_{4})$ before 
integrating over energies.
In analogy with eq.~(\ref{EQU:utgpkt}), we introduce
\beq
\dd^8\tilde{\Gamma}
= \dd^8\Gamma (\omega_{1}-\omega_{2})(\omega_{3}-\omega_{4}) ,
\nn
\eeq
and integrate analytically over kinematic variables to obtain the
distribution 
\beqa
 \frac{\dd^3\tilde{\Gamma} }{\dd\phi\: \dd s_1 \dd s_2}
& =      &\frac{\sqrt{2}}{288 (4\pi)^6}\, N_1 N_2 \, \frac{G_{\rm F}}{m^5}
\lambda^{3/2}\left(m^2,s_1,s_2\right)\, D(s_1,s_2) \\
& & \times 
\left[ 2s_1 s_2\,\sin 2 \chi_{1} \sin 2 \chi_{2} \rho^2
- \left(\frac{3 \pi}{16}\right)^2
\sqrt{s_1 s_2}\, (m^2-s_1-s_2)\,\rho\,\cos (\phi - \delta) \right]. \nn 
\label{EQU:no31}
\eeqa
In the narrow-width approximation we obtain
\beq
\frac{2 \pi}{\tilde{\Gamma}}
\:\frac{\dd\tilde{\Gamma}}{\dd\phi} 
 =   
1 + \frac{\kappa\left(m\right)}{\rho } \cos (\phi-\delta),
\label{EQU:Dl7}
\eeq
with
\beq
\kappa(m) =
-\frac{1}{\sin 2 \chi_{1} \sin 2 \chi_{2}}
\left(\frac{3 \pi}{16}\right)^2 \frac{2-\mu}{\mu}
=\frac{1}{\left(4 \sin 2 \chi_{1} \sin 2 \chi_{2}\right)^{2}} 
\frac{\beta(m)}{\alpha(m)},
\label{EQU:Dkappa}
\eeq
which is independent of $\eta$.
As in eqs.~(\ref{EQU:nr1})--(\ref{EQU:nr2}) we introduce the
following measures of asymmetry: 
\beq
\left.
\begin{array}{c}
C\left(m\right) \\
C^{\prime}\left(m\right)
\end{array}
\right\}
\equiv 
\frac{1}{\pi} \int_{0}^{2\pi}\dd\phi 
\left\{
\begin{array}{c}
\cos \phi \\
\sin \phi
\end{array}
\right.
\left(\frac{2 \pi}{\tilde{\Gamma}}\:\frac{\dd\tilde{\Gamma}}{\dd\phi}\right) = 
\left(\frac{\kappa\left(m\right)}{\rho}\right)
\left\{
\begin{array}{c}
\cos \delta \\
\sin \delta
\end{array}
\right., 
\label{EQU:nr02} 
\eeq
so that 
\beq
\frac{2 \pi}{\tilde{\Gamma}}
\:\frac{\dd\tilde{\Gamma}}{\dd\phi} 
 =   
1 + C\left(m\right) \cos \phi
+ C^{\prime}\left(m\right) \sin \phi.
\label{EQU:Dlk7}
\eeq
Consequently
\beq
\tan \delta  = \frac{C^{\prime}\left(m\right)}{C\left(m\right)}.
\nn
\eeq
Hence, this set of observables provides yet another way of measuring 
the $CP$ violating phase $\delta$. 
Although the ratio of these energy-weighted
observables coincides with, or is trivially related to the previous ones, 
the possibility of demonstrating a presence of $CP$ violation has increased
significantly. The ratio
\beq
\frac{C^{\prime}\left(m\right)}{B^{\prime}\left(m\right)} = \frac{1}{\left(4
\sin 2 \chi_{1} \sin 2 \chi_{2} \right)^{2} \alpha\left(m\right) \rho^2}
\label{EQU:forh2}
\eeq
is given in fig.~\ref{plfun6} for $\eta=10^{-3},
10^{-1}$, and $1$ in the
cases of $H \rightarrow W^{+}W^{-}\rightarrow 4 f$ and
$H \rightarrow ZZ\rightarrow 4 l$. 
We see that the energy-weighted
$CP$-violating observable $C^{\prime}$ is
a much more sensitive probe for
establishing $CP$ violation than the former ones. Moreover, the amplitude 
function $\kappa(m)/\rho$ of (\ref{EQU:Dl7}) is given in 
fig.~\ref{plfun4} for the cases 
$H \rightarrow W^{+}W^{-}\rightarrow 4 f$ and 
$H \rightarrow ZZ\rightarrow 4 l$. 
We observe that this amplitude is comparable to,
or much bigger than unity for arbitrary values of $\eta$ 
and for Higgs decay to any observable four-fermion final state.
In addition, for $V=Z\rightarrow 2l$ the $\sin 2\chi$-factors 
provide an enhancement in the energy-weighted amplitude function, and
particularly in the ratio $C^{\prime}/B^{\prime}$. 
It is encouraging that this enhancement occurs for the so-called
``gold--plated mode'' $H \rightarrow ZZ\rightarrow
e^{+}e^{-}\mu^{+}\mu^{-}$ \cite{HHG}.

In conclusion, we have demonstrated that, 
for an arbitrary amount of $CP$ violation in the $HVV$
coupling, the distribution (\ref{EQU:intro1}) 
has the compact and transparent form (\ref{EQU:Dl5}). 
As compared with the observables $A^{\prime}$ and $B^{\prime}$,
presence of $CP$ violation would be much more easily measured 
by using the energy-weighted observable $C^{\prime}$. 
Finally, having established $CP$ violation, the actual
strength may be determined from measurements of the observables $C$ and 
$C^{\prime}$.


This research has been supported by the Research Council of Norway.
                
\renewcommand{\arraystretch}{1.5}
\clearpage
\centerline{\bf Figure captions}

\vskip 15pt
\def\fig#1#2{\hangindent=.65truein \noindent \hbox to .65truein{Fig.\ #1.
\hfil}#2\vskip 2pt}

\fig1{The amplitude functions $\alpha(m)\rho^2$ and
$-\beta(m)\rho$ of (\ref{EQU:Dl5}) in the case 
$H \rightarrow W^{+}W^{-}\rightarrow 4 f$, for a Higgs of mass $m$, 
and for $\eta=1$, $10^{-1}$ and $10^{-3}$.}

\fig2{The amplitude functions $\alpha(m)\rho^2$ and
$-\beta(m)\rho$ of (\ref{EQU:Dl5}) in the case 
$H \rightarrow ZZ \rightarrow 4 l$, for a Higgs of mass $m$ and for 
$\eta=1$, $10^{-1}$ and $10^{-3}$.}

\fig3{The angle $\delta$ (in degrees) for a Higgs particle
of mass $m$, for $\eta=1$, $10^{-1}$, $10^{-2}$, and $10^{-3}$.}
        
\fig4{The ratio $C^{\prime}\left(m\right)/B^{\prime}\left(m\right)$ for a Higgs
of mass $m$ in the cases of 
$H \rightarrow W^{+}W^{-} \rightarrow 4 f$ and
$H \rightarrow ZZ \rightarrow 4 l$, for $\eta=10^{-3}$, $10^{-1}$, and $1$.}

\fig5{The amplitude function $-\kappa(m)/\rho$ 
in the cases 
$H \rightarrow W^{+}W^{-}\rightarrow 4 f$ and 
$H \rightarrow ZZ \rightarrow 4l$, for a Higgs of mass $m$, 
and for $\eta=1$, $10^{-1}$, and $10^{-3}$.}


\clearpage

\clearpage

\clearpage

\clearpage

\clearpage

\end{document}